\documentclass[letter]{aa}     

\usepackage{graphicx}
\usepackage{txfonts}
\usepackage{lipsum}
\usepackage{subcaption}         
\usepackage{lscape}             
\usepackage{placeins}           
                                
\usepackage[hidelinks]{hyperref}
\usepackage{xcolor}
\hypersetup{
    colorlinks = true, 
    linkcolor={red!50!black},
    citecolor={blue!50!black},
    urlcolor={blue!80!black}
}
\begin{document}
%
%
   \title{Pulsation-driven helium transport as a potential source of the\\ 
          Blazhko effect\thanks{This work is dedicated to the living memory  
	  of my mentors and friends Wojtek Dziembowski and Robert Buchler.}}
%
%

%

   \author{Geza Kovacs\inst{1}
        }
   \institute{Konkoly Observatory, Research Center for Astronomy and Earth Sciences  
              of HUN-REN \\ 
              Budapest, 1121 Konkoly Thege ut. 15-17, Hungary\\
              \email{kovacs@konkoly.hu}}

   \date{Received 22-03-2026 / Accepted DD-MM-2026}
 
%
%

%
  \abstract 
{We present a highly simplified nonlinear hydrodynamical model to emulate 
the main observed features of amplitude modulation (commonly known as 
Blazhko effect) in RR Lyrae stars. The model is based on the assumption 
that the periodic flow generated by the pulsation carries surplus helium 
in the ionization zones He I and II. Once this extra helium reaches 
a critical amount, a Rayleigh-Taylor-type instability leads to a back-flow 
of the surplus helium and the process starts over again, due to the 
continuing effect of pumping helium upward by the pulsation. This periodic 
variation of helium leads to various efficiency of radiation flux 
blocking in the helium ionization zone that shows up as a long-term periodic 
variation of the pulsation amplitude.}


   \maketitle
\nolinenumbers

%
%
\section{Introduction}
\label{sect:intro}
\begin{figure}[h!]
   \centering
   \includegraphics[width=0.40\textwidth]{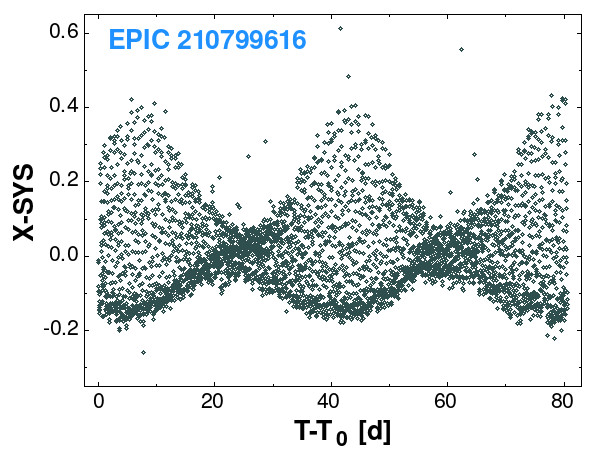}
      \caption{An example of the highly-modulated RR Lyrae stars.
               The data are from the Kepler K2 mission. The 
	       vertical axis is the normalized stellar flux after 
	       filtering out instrumental systematics.}
         \label{high_BL}
   \end{figure}

Fundamental mode RR Lyrae stars are unique among the classical 
large amplitude pulsators due to their substantial amplitude 
modulations \citep{blazhko1907,shapley1916}. The phenomenon is 
very frequent among the fundamental mode RR Lyrae stars 
\citep[at least $75$\% -- see][]{kovacs2025}. The ratio of the 
maximum and minimum amplitudes is indeed considerable, sometimes 
it may reach several factors (see Fig.~\ref{high_BL} for a 
spectacular example). There are also stars with very low modulation 
amplitudes, more difficult to detect, but, starting from the 
early days of astronomical photometry, the phenomenon was always 
within reach by the available telescopes.  

%
%

A large amount of work has been invested in the empirical study of 
the Blazhko phenomenon during the past $\sim 100$ years, looking for 
some empirical clue of the physical origin of the amplitude modulation. 
Understandably, this was the only way to gather information about 
these fascinating objects, because no theory was available with the 
potential of hydrodynamical modeling to aid a useful comparison of 
the model predictions with the observations. Multitude of ideas 
came up during the past $\sim 70$ years, but none of them proved 
to be workable and physically sound (without completeness, we refer 
to the reviews of \citealt{stellingwerf1976}, \citealt{kovacs2009}, 
and \citealt{kollath2018} for the details of the past theoretical 
efforts). 

This work has been largely motivated by the size of the modulation. 
It is clear that to produce a variation like the one displayed in 
Fig.~\ref{high_BL}, we need to think about something that is vital 
for the excitation mechanism. Obviously, we cannot resort to non-radial 
modes, because of the huge distortion required by the observed 
destruction and subsequent build-up of the large amplitude radial 
pulsation. Also, in most cases, the pulsation remains monomode, i.e., 
the modulation is not a by-product of some sort of double-mode 
pulsation, but rather a genuine modulation of the single-mode nonlinear 
pulsation.   

\begin{figure}[h!]
   \centering
   \includegraphics[width=0.45\textwidth]{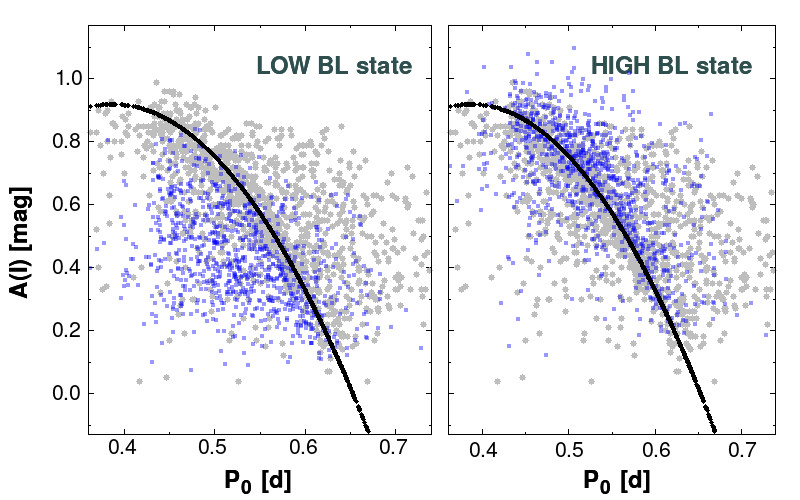}
      \caption{Bailey plots (pulsation period vs total amplitude) 
               for the Galactic Bulge RR Lyrae star 
	       \citep{colligne2006, skarka2020}. Gray dots are for 
	       the fundamental mode monoperiodic, blue dots for the 
	       Blazhko stars. Low and high BL states, respectively, 
	       denote low and high amplitude episodes during the 
	       Blazhko cycle. The continuous line is an eye-fitted 
	       parabola to the ridge of the gray points.}
         \label{bailey_bulge}
   \end{figure}

Here we remain in the realm of purely radial pulsation and utilize 
a simple particle transport (toy\footnote{That is, the transport 
mechanism itself is not modeled, but assumed to act under certain 
physical conditions.}) model and aim at the capability of reproducing 
all basic properties of the Blazhko phenomenon. For a brief overview, 
we list these properties in Table~\ref{BL_prop}. We add the following 
notes to the items shown. Item \#1 indicates that the pulsation can 
be nearly completely seized in some cases. The reliability of the 
long period modulations comes mostly from the ground-based long-term 
photometric surveys. In more detail, \#3  says that an overwhelming 
number of the Blazhko stars show an anti-correlation in the 
variation of the pulsation maxima and minima \citep[see][]{skarka2020}. 
We found this property a critically important one in constraining 
the transport model presented in this paper. Property \#4 does not 
say that all Fourier side peaks should always be asymmetric, but that 
this property is very common \citep[see][]{alcock2003}. \cite{szeidl1988} 
has shown (though on a rather limited sample, available at that time) 
that the high amplitude state of the Galactic field and M3 cluster 
variables fit the monomode ridge of the Bailey diagram, and all the 
low amplitude states of the Blazhko variables are under that ridge. 
We repeated his work on a sample of the Galactic Bulge RR Lyrae stars 
and found this property strongly supported by these recent data 
(see Fig.~\ref{bailey_bulge}).  

%
%
\begin{table}[h]
\begin{flushleft}
\caption{Overall properties of fundamental mode Blazhko stars.}
\label{BL_prop}
\scalebox{1.0}{
\begin{tabular}{ll}
\hline
Property               &  Value \\
\hline\hline
{\em 1. Modulation amplitude:}  &  Up to the pulsation amplitude \\
{\em 2. Modulation period:}     &  $10-1500$ days  \\
{\em 3. Upper and lower}        &   \\
{\em \quad modulation envelopes:}  &  Anti-correlated \\
{\em 4. Fourier side peaks:}    &  Asymmetrical \\
{\em 5. Bailey diagram:}        &  High amplitude Blazhko state \\
                             &  is close the monoperiodic ridge \\
\hline 
\end{tabular}}
\end{flushleft}
\end{table}
%

%
%
%

%
%
\section{The model}
\label{sect:model}
\begin{figure}[t!]
   \centering
   \includegraphics[trim={0 0pt 0 0}, width=0.40\textwidth, clip=true]{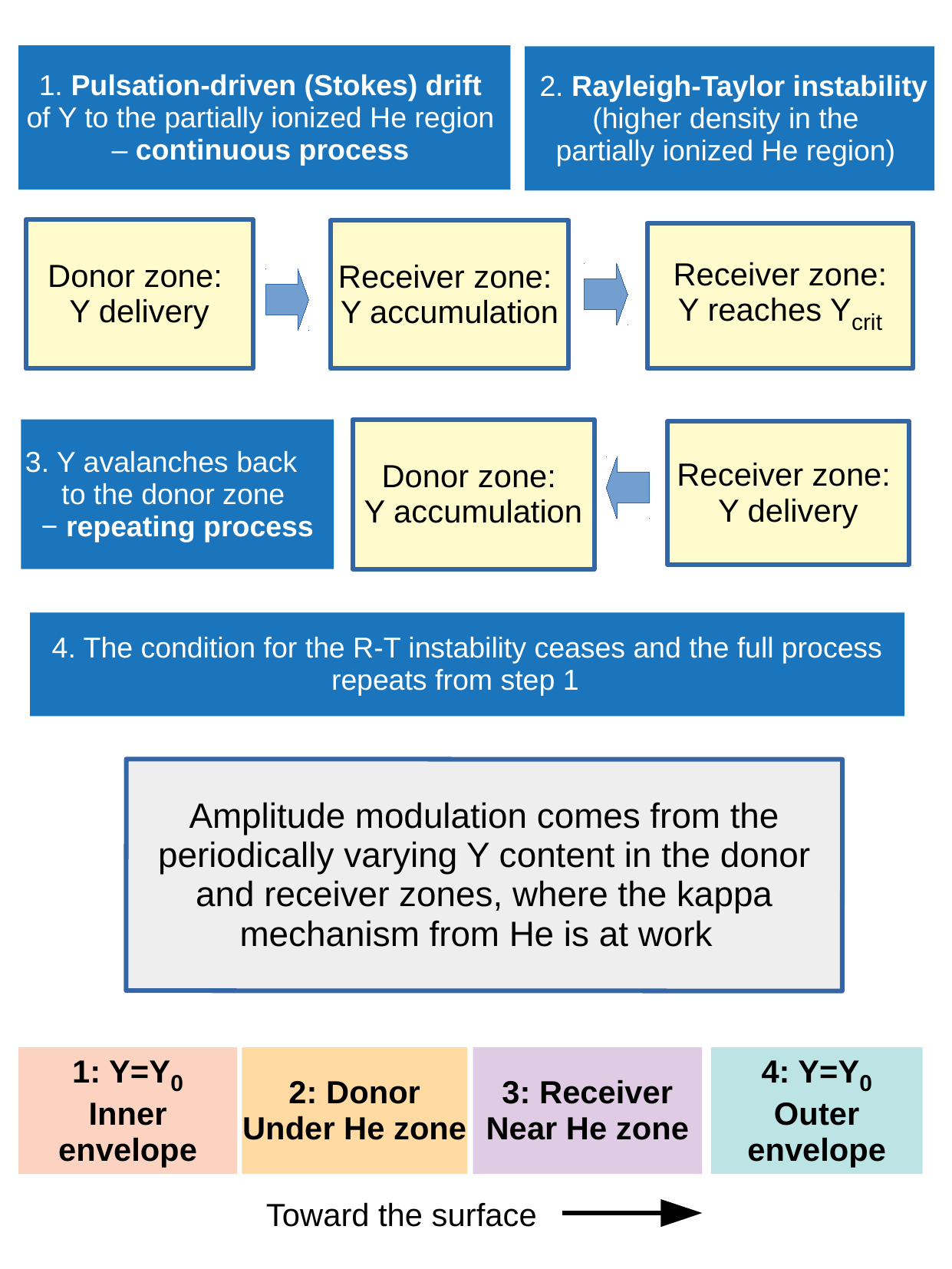}
      \caption{The main ingredients of the helium (Y) transport model 
               for the periodic modulation of the pulsation amplitude.}
         \label{flow_chart}
   \end{figure}

The basic steps of the dynamical transport process suggested are 
depicted in Fig.~\ref{flow_chart}. The core of the idea is the 
cyclic variation of the helium, that leads to changes in the 
opacity in the helium ionization zone\footnote{For simplicity, we 
refer to the He I and II ionization zones in a singular form.} 
and thereby causing variation in the amplitude of the pulsation. 
The more regular motion generated by the pulsation below the helium 
ionization zone  allows the delivery of the helium into higher 
levels of the envelope, where the temperature is lower. This happens 
most effectively during the expansion phase, which is the most 
violent part of the pulsation cycle for RR Lyrae stars, pulsating 
in the fundamental mode. The transport mechanism is likely similar 
to the well-known Stokes 
drift\footnote{\url{https://en.wikipedia.org/wiki/Stokes_drift}}, 
often encountered in (quasi-)periodic fluid flows, such as ocean 
waves. Both hydrogen and helium are transported, but hydrogen 
overabundance becomes relevant only in the upper layers, where the 
flow is far less regular and no significant accumulation of hydrogen 
is expected there. During the pulsation, in the descending light phase, 
when the material flows downward, the element transport reverses. 
We assume that the net outcome of the particle motion is the win 
of the upward drift, due to the much stronger forces (including 
shocks) acting on the particles in the expansion phase (in spite 
of its short duration with respect to that of the down-drift 
during the free-fall phase). If the pulsation is indeed capable 
of delivering enough helium in the part of the stellar interior 
where the motion is still close to laminar, this will lead to a 
density inversion and the material starts to fall back once the 
density reaches a certain critical value -- this is the familiar 
Rayleigh-Taylor 
instability.\footnote{\url{https://en.wikipedia.org/wiki/Rayleigh-Taylor_instability}}
While the pulsation drives helium upward on the time scale of the 
pulsation, the development of a sufficient helium gradient takes 
time. Eventually, this leads to a periodic process on a long time 
scale, involving the increase and the subsequent partial depletion 
of helium in the ionization zone of this element. 

%
%

\begin{figure}[h!]
   \centering
   \includegraphics[width=0.40\textwidth]{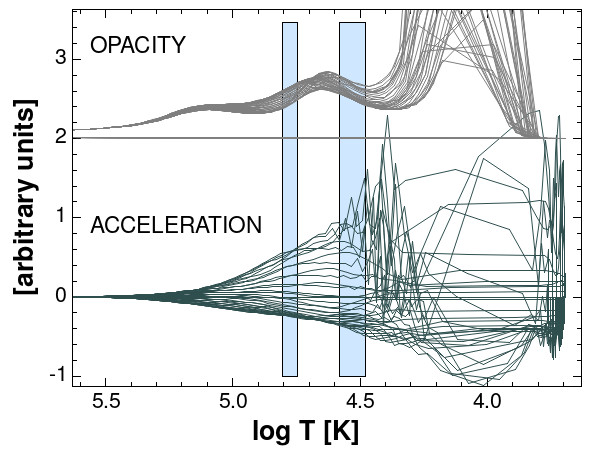}
      \caption{Variation of the opacity and the acceleration 
               through the fundamental single mode limit cycle 
	       in the interior of the chemically homogeneous 
	       model. Each line corresponds to a snapshot during 
	       the pulsation cycle. The two vertical stripes show 
	       the temperature ranges covered by the limits 
	       of zone 3, as these limits change during the 
	       pulsation and in the Blazhko cycle. The pulsation 
	       is more dynamical above the upper bound of zone 3, 
	       toward the stellar surface at lower temperatures, 
	       including the hydrogen ionization front.}
         \label{opac_acc}
   \end{figure}

In the lower part of Fig.~\ref{flow_chart} we symbolically show the 
four regions of the stellar interior that are relevant for the 
pulsation and for the simple transport model briefly described above. 
At the innermost and outermost zones\footnote{We use the word ``zone'' 
for the regions relevant from the point of view of the helium transport, 
and ``mass shell'' for the Lagrangean description of the pulsation as 
actually used in the hydrodynamical code.} 
we assume that the chemical composition is constant, either because 
of the decreasing efficiency of the pulsation (zone 1) or, because 
of the the more turbulent motion (zone 4). Zones 2 and 3 are the 
transport zones, where the periodic exchange of the helium occurs 
due to the relative laminarity of the flow (a few $10^4$~K beneath 
the region of the hydrogen ionization -- see Fig.~\ref{opac_acc}). 
The basic assumption is the sufficient tranquility of zone 3 that 
enables the helium drifted upward from zone 2 to stay in this region 
for some period of time (of the order of the Blazhko period). 
This condition may be violated if convective overshooting becomes 
too effective in dispersing He from zone 3. 
   
%
%
%

%
%
\subsection{The hydrodynamical implementation}
\label{sect:hydro}
We use the same, purely radiative, nonlinear radial pulsation, 
direct time integration code as in our earlier studies. The code 
is a slightly modified version of the one used in the seminal 
paper of \cite{stellingwerf1975}. It is a Lagrangean code with 
gray radiative transfer. In the present implementation we use 
$100$ mass shells with $15$ shells of equal mass from the surface 
down to the hydrogen ionization zone at $11000$~K (as set in 
the static model). From here on, the masses of the shells 
are geometrically increasing down to $7\times10^6$~K. This is 
an envelope model, comprising some $94$\% of the stellar radius 
and $10$\% of the total mass. The stellar parameters are typical 
for an RR Lyrae star, namely: 
$M/M_{\odot}=0.65$, $L_{\odot}=40$, $T_{\rm eff}=6500$~K, 
$X=0.76$, $Y=0.239$, $Z=0.001$. The fundamental period is 
$0.488$~d. To avoid non-essential complications, we use 
Stellingwerf's analytical formula to compute the Rosseland 
mean opacity. 

At each step of the integration of the model, we modify the 
local amount (i.e., particle number) of helium, depending 
on the zone number and local conditions.\footnote{Note that 
when the local (i.e., per mass shell) composition is fed into 
the equation-of-state and opacity routines, we have to renormalize 
the particle number to satisfy the $X+Y+Z=1$ condition.} 
The scheme should, of course, preserve the total number of 
particles (i.e., the overall $X$, $Y$, $Z$ abundance ratios) 
throughout the entire simulation. 

In our model, only zone 2 delivers $Y$ due to 
pulsation.\footnote{Zone 3 may also deliver $Y$ into zone 2 
during the down-drift phase, but we count the back flow 
(at least for this study) only as an efficiency decreasing 
factor for the up-drift from zone 2.} In the up-drift phase 
(expansion) $Y$ is shovelled into zone 3. The down-drift 
phase (recession) in zone 2 is constructed in a way that lets $Y$ 
transport die out in deeper mass shells, due to the low amplitude of 
the pulsation.  

In zone 2, the amount of $Y$ transferred from mass shell $i$ to 
$i+1$ is determined by the local conditions, and can be expressed 
as follows. The acceleration in the $i-$th shell is denoted by 
$a_i$. If $a_i\leq 0$, then no transport occurs (see above). 
For $a_i>0$, we have $Y_i = Y_i - \Delta Y$ and 
$Y_{i+1} = Y_{i+1} + \Delta Y$, where 
$\Delta Y= D_u a_i dt^2 / (r_i-r_{i-1})$, with $dt$ time step 
and $r_i$ shell outer radius. The time independent factor $D_u$ 
represents the efficiency of the element transport through the 
expanding phase of the pulsation. 

At each time step, the code checks whether the ratio of $Y$ in zone 
3 to zone 2 exceeds a certain critical number given by the parameter 
$R_2$. If it happens, the avalanche of $Y$ starts from zone 3 back 
to zone 2, and it continues until the ratio remains greater than 
some value, given by the minimum $Y$ ratio parameter, $R_1$. The 
amount of $Y$ delivered during one time step is given by another 
constant parameter, denoted here as $D_a$. 

Except for the high temperature limit of zone 2, the mass shell 
indices of the zone limits are determined from the temporal local 
temperatures. The innermost shell index $i_1$ for zone 2 is set 
equal to 20 (corresponding to $T\sim 7\times 10^5$~K). The temperature 
at the outermost border of this zone is fixed at $T_2=6\times 10^4$~K, 
and the corresponding zone index $i_2$ is chosen as the index of 
the shell nearest to this temperature. The outer boundary of zone 3 
is simply $i_3=i_2+\Delta i$, where $\Delta i=10$. With this choice, 
the helium ionization zone is largely within zone 3. We ended up 
with this setting after a considerable amount of experimentation 
to reach robust stability in the modulated models and, in particular, 
approaching to the desired observed properties. Further discussion 
of the effects of transport zone selection is given in Appendix~\ref{app_A}.   

Another very important part of the transport process is the in-zone 
flattening and linear ramping of the abundances at the zones 
limits. The abundances are ramped to the static values at the 
zone borders. Without ramping, both the robustness and observational 
compatibility (in particular, property 3 of Table~\ref{BL_prop}) of 
the amplitude modulation and the stability of the numerical scheme 
are jeopardized. 

In summary, the above model uses six parameters. The up-drift and 
avalanche efficiencies $D_u$ and $D_a$, the critical $Y$ avalanche 
ratios $R_1$ and $R_2$, the temperature at the border of the two 
transport zones and the thickness of zone 3 ($T_2$ and $\Delta i$, 
respectively). The values of these parameters can be adjusted by 
observing the outcome of the hydrodynamical runs. For example, the 
relative values of $D_u$ and $D_a$ determine the time spent in the 
amplitude increasing/decreasing phases. Obviously, they also have 
an effect on the length of the Blazhko period, together with $R_1$ 
and $R_2$, that control the amplitude of the modulation. 
Furthermore, we found that the condition $R_1 \geq 1$ should be 
satisfied if property 5 in Table~\ref{BL_prop} is required to be 
observed.   

%
%
\section{Examples}
\label{sect:examples}
We illustrate the performance of the $Y$ transport model resulting in 
a moderate (Fig.~\ref{hydro1}) and in a large amplitude modulations 
(Fig.~\ref{hydro2}). For a relative amplitude modulation of 
$\sim 10-30$\% the required change in $Y$ in the helium ionization region   
(zone 3) is only a few hundredths. A factor of two, or even larger 
amplitude changes require considerable amount of change in $Y$, 
typically $\pm (0.07-0.10)$. Because zone 3 is confined only in ten mass 
shells, the change in the more extended donor zone (zone 2) is considerably 
smaller, i.e., $\pm (0.015-0.030)$. Figure~\ref{hydro1} shows that the 
change in $Y$ in zone 3 is anticorrelated with that of the amplitude.
We think that in the high-amplitude phase this effect comes from the 
lower concentration of the helium in the ionization zone, that leads 
to higher photon flux. This is then reflected in the higher pulsation 
amplitudes, and also, in the overall brightening of the star during 
the high amplitude phase of the Blazhko cycle. This phenomenon is present 
both in the observed stars and in the models. The brightening appears 
in the Fourier spectrum (lower panel of Fig.~\ref{hydro2}), as a low 
frequency peak at the modulation frequency. We also note that the 
sporadic reports on helium emission in Blazhko stars 
\citep[i.e.,][]{gillet2013} may also support the above explanation, 
although the emission is also observable in monomode RR Lyrae stars 
\citep[][]{preston2022}. Figure~\ref{hydro2} also shows that the side 
lobes are asymmetrical, in consonance with the observations 
\citep[i.e.,][]{alcock2003}. In spite of this comforting similarity, 
we note that (at least within the extent of this exploratory study) 
frequency patterns with opposite asymmetricity were faraway from an 
easy access.   

%
%
   \begin{figure}[h!]
   \centering
   \includegraphics[width=0.40\textwidth]{hydro_pm0.jpg}
      \caption{Example of a moderate amplitude modulation with the 
               period of $38$~d. The associated variations of $Y$ 
	       in zones 2 and 3 are shown in the lower part of the 
	       figure. The ratio of the pulsation amplitudes in the 
	       high- vs low-amplitude phases of the modulation is $1.2$. 
	       Note that this modulation is reached by a mere $\pm 0.02$  
	       variation of Y in the helium ionization zone.}
         \label{hydro1}
   \end{figure}
%

%
%
   \begin{figure}[h!]
   \centering
   \includegraphics[width=0.40\textwidth]{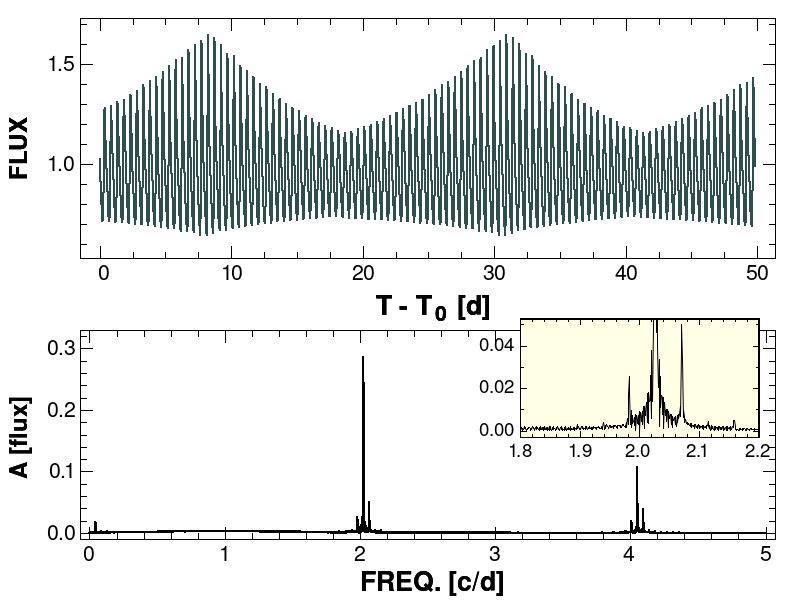}
      \caption{{\em Upper panel:} as in Fig.~\ref{hydro1}, for the 
               a model with the same stellar parameters, but larger 
	       Y transport efficiencies ($D_u$ and $D_a$) and 
	       avalanche parameter $R_2$. The pulsation amplitude 
	       ratio in the high and low states is $2.39$. The 
	       modulation period is $23$~d.     
	       {\em Lower panel:} Fourier spectrum of a long stretch 
	        of the hydrodynamical run, showing the asymmetrical 
		modulation components. The inset zoomes in the 
		neighborhood of the fundamental frequency component.}
         \label{hydro2}
   \end{figure}
%

%
%
\section{Conclusions}
\label{sect:summary}
We presented the idea of periodic helium transport as the source of 
the long-term amplitude modulation observed in RR Lyrae stars. The 
modulation period corresponds to the cycle of the steady upward drift 
of the helium from the region below the helium ionization zone, its 
accumulation in the ionization zone, and then the subsequent avalanche 
of it back to the feeding zone. The varying helium blocks some 
fraction of the outgoing stellar flux, and then releases it, 
as exhibited by the larger pulsation amplitude and by the slightly 
increased overall stellar luminosity in the high-amplitude state. 

The idea was implemented in a nonlinear hydrodynamical code through 
an algorithm that let helium be transported and then down-drifted 
when certain physical conditions are satisfied. Although this treatment 
of element transport is not based on first principles (i.e., 
not the result of ab initio heterogeneous particle hydrodynamics), 
it is not an external modulation injected in a standard pulsation 
model. The modulation is the outcome of the response of the system 
to the changing chemical composition, induced by the (not entirely 
unsound) physical conditions under which the element transport can 
take place. In this sense, the model presented in this paper is 
the first autonomous nonlinear RR Lyrae model capable of producing 
the basic observed properties of the Blazhko stars. It is also 
remarkable, that during the extensive parameter testing of the 
various helium transport ideas, only the one presented in this 
paper came out as a successful scheme, allowing parametric modeling 
many characteristics of the Blazhko modulation. Clearly, only 
detailed hydrodynamical simulations with realistic chemical transport 
models will be able to support or reject the instrumental nature of 
pulsation-driven element transport claimed here for the explanation 
of the Blazhko effect.

\begin{acknowledgements}
We are grateful to Santi Cassisi and Noam Soker for the earlier 
cordial correspondence on topics related to some ``other'' failed 
ideas on the subject of this paper.
%
This paper includes data collected by the Kepler mission and obtained 
from the MAST data archive at the Space Telescope Science Institute 
(STScI). Funding for the Kepler mission is provided by the NASA 
Science Mission Directorate. STScI is operated by the Association 
of Universities for Research in Astronomy, Inc., under NASA contract 
NAS 5-201326555.
\end{acknowledgements}

%
%
%

%
%
%
%
\begin{appendix}
\section{Effect of transport zone limits}
\label{app_A}
In part for the request of the referee, here we present tests addressing 
the following two questions.
 
\noindent 
a.) How sensitive is the modulation to the width of the receiver zone?  
-- Knowing that this zone overlaps with the convectively unstable He 
ionization region, leading to mixing and smearing.  

\noindent 
b.) What are the conditions for the transport zone limits to maintain 
anti-correlation between the upper and lower envelopes of the modulation? 
-- An important observational constraint for the model. 

We tested the same RR~Lyrae model as the one detailed in the main part of 
the paper. The transport model was also the same, except that here the 
position of the upper end (the low temperature end) of the receiver zone 
(zone 3) was given by the temperature parameter $T_3$ rather than the 
zone number $i_3$, which is determined by the fixed zone number difference 
between the lower and upper bounds of zone 3. The two settings are very 
similar, except that the one used here is more physical, because it 
refers to temperatures, insted of zone numbers. 

For a more transparent model description, here we introduced the following 
scaling. For the pulsation-driven $Y$ transport, the efficiency coefficient 
was set as follows: $D_{u} = f_{u}D_{u}^0$, where $D_{u}^0=0.11$, and all 
related quantities (acceleration, time step, radii -- see 
Sect.~\ref{sect:hydro}) are in CGS units. The avelanche efficiency was  
formulated similarly: $D_{a} = f_{a}D_{a}^0$, where $D_{a}^0=0.000011$. 
The avelanche condition (see Sect.~\ref{sect:hydro}) was set for all 
models in the same way, namely $R_1=1.0$, $R_2=1.3$. With the fundamental 
period of $P_0\sim 0.493$~d, the time step was chosen as $P_0/443$. 
Although all models reached steady states within one thousand fundamental 
periods, to check stability, the integration continued further for 
several thousand periods.  

%
%
\begin{table}[h!]
\begin{flushleft}
\caption{Transport parameter for the modulation tests.}
\label{tz_sets}
\scalebox{0.96}{
\begin{tabular}{cccccccccc}
\hline
 Run  & $T_2$ &  $T_3$ &  $\Delta T$  & $f_{u}$ & $f_{a}$ & $P_m$ & $A_m$ & $EC$ & $\langle A\rangle$ \\
\hline\hline
  00  &  2.0  &  0.8  &  1.2  &  1.0  &  3.0  &  -   &   -    &  -   &    -  \\
  01  &  2.5  &  0.8  &  1.7  &  0.5  &  1.0  &  -   &   -    &  -   &    -  \\
  02  &  3.0  &  1.0  &  2.0  &  0.1  &  0.3  &  47  &  0.04  &  PC  &  0.77 \\ 
  03  &  4.0  &  1.0  &  3.0  &  0.5  &  0.4  &  41  &  0.13  &  PC  &  0.77 \\ 
  04  &  5.0  &  1.0  &  4.0  &  1.0  &  0.5  &  32  &  0.26  &  PC  &  0.75 \\  
  05  &  5.5  &  1.5  &  4.0  &  1.5  &  0.5  &  32  &  0.32  &  NC  &  0.76 \\ 
  06  &  6.0  &  3.0  &  3.0  &  1.5  &  0.7  &  30  &  0.35  &  NC  &  0.71 \\ 
  07  &  6.5  &  2.5  &  4.0  &  1.7  &  0.5  &  37  &  0.33  &  NC  &  0.72 \\ 
  08  &  7.0  &  2.0  &  5.0  &  2.0  &  0.4  &  41  &  0.33  &  NC  &  0.74 \\ 
  09  &  7.5  &  1.5  &  6.0  &  3.0  &  0.4  &  37  &  0.32  &  PC  &  0.78 \\ 
  10  &  8.0  &  1.0  &  7.0  &  3.5  &  0.4  &  37  &  0.30  &  PC  &  0.79 \\ 
\hline 
\end{tabular}}
\tablefoot{The temperature limits of the receiver zone $T_2$, $T_3$ 
           and their difference $\Delta T$ are given in $10^4$~K. The 
	   modulation period $P_m$ is in days, the modulation and the 
	   average amplitudes ($A_m$ and $\langle A\rangle$) are in 
	   relative fluxes (i.e., dimensionless). The pulsation-driven 
	   drift and avalanche factors $f_{u}$ and $f_{a}$ are also 
	   dimensionless. The envelope correlation $EC$ is characterized 
	   by the acronyms PC and NC (for positive and negative 
	   correlations, respectively).}
\end{flushleft}
\end{table}

Table \ref{tz_sets} shows the main parameters of the runs. The transport 
parameters are chosen to reach roughly equal durations for the amplitude 
increase and decrease phases and retain the modulation periods in a 
relatively confined interval. For the first two runs the models switched 
to the first overtone. These models basically avoid the main part of the 
He ionization zone, and deliver $Y$ to the upper envelope, including the 
hydrogen ionization zone. 

Models 02-04 show amplitude modulation, in spite of the partial coverage 
of the He ionization region by zone 3. The lack of the flux-blocking 
effect of the He ionization zone is especially spectacular for model 02. 
The amplitude modulations are small in the absolute sense, because almost 
independently of the position of zone 3, the average amplitudes of the 
models are very similar (see column 10 of Table~\ref{tz_sets}). 

For models 08-10, zone 3 covers wide temperature ranges, from the deeper 
interior (below the He ionization region) to the H ionization region. 
Based on these results, it seems quite secure to say -- as an answer on 
question (a) -- that even for considerable convective mixing and 
overshooting, leading to the extension of zone 3, amplitude modulation 
via the suggested transport mechanism works efficiently. 

\begin{figure}[h]
\centering
\includegraphics[width=0.48\textwidth]{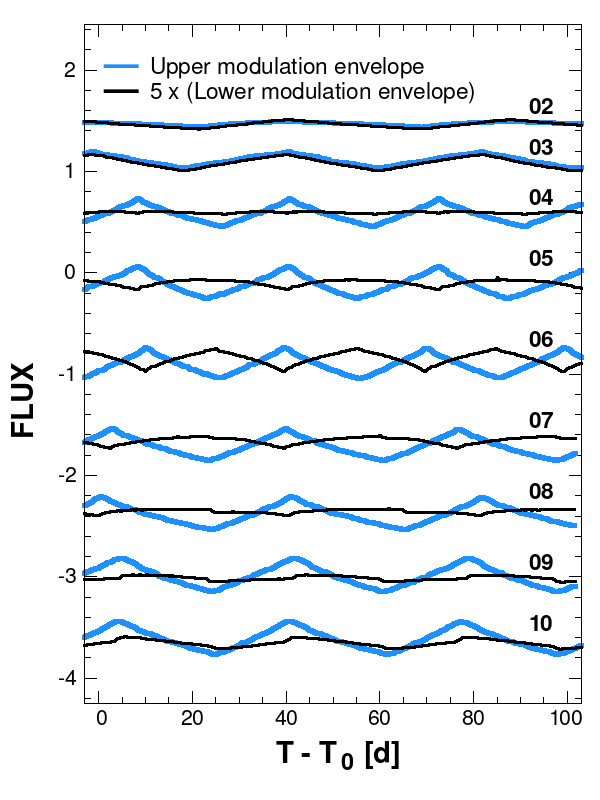}
\caption{Variation of the upper and lower envelopes of the pulsation 
         amplitude modulation for different transport zone settings 
	 (see Table~\ref{tz_sets} and the code numbers on the right). 
	 All envelopes are plotted on the same scale, but shifted 
	 properly for better visibility. For the same reason, the 
	 amplitudes of the lower envelope curves have been multiplied 
	 by a factor of five. For matching the moments of one of the 
	 maximum modulations at $T-T_0=40$, the initial epoch ($T_0$) 
	 is different for each run. See text for the discussion 
	 of the trends follow from this plot.    
	 } 
\label{anti-corr}
\end{figure}

In spite of the overall flexibility of the occurrence of modulation 
with respect to the position and width of zone 3, satisfying the 
anti-correlation criterion for the modulation envelopes (requirement (b)), 
considerably constrains the available temperature limits for zone 3. 
Table~\ref{tz_sets} and Fig.~\ref{anti-corr} show that all preferable 
models (05-06-07) have limited temperature ranges for zone 3 and all 
are right at the He ionization region. This result implies that if 
periodic He transport is indeed the cause of the Blazhko effect, then 
the convective zone in the He ionization region must be well-confined 
to this region.

%
%

\end{appendix}
\end{document}